\documentclass[prl,amsmath,reprint]{revtex4-1}
\usepackage{color, graphicx}
\usepackage[version=3]{mhchem} 
\usepackage{subfig}
\usepackage{bm} 
\usepackage{amsmath}

\begin{document}
\title{Microscopic origin of entropy-driven polymorphism in hybrid organic-inorganic perovskite materials}

\author{Keith T. Butler}
\email{k.t.butler@bath.ac.uk}
\affiliation{Department of Chemistry, University of Bath, Claverton Down, Bath BA2 7AY, UK}
\author{Gregor Kieslich}
\affiliation{Department of Materials Science and Metallurgy, University of Cambridge, Cambridge CB3 0FS, UK}
\email{gk354@cam.ac.uk}

\author{Katrine Svane}
\affiliation{Department of Chemistry, University of Bath, Claverton Down, Bath BA2 7AY, UK}

\author{Anthony K. Cheetham}
\affiliation{Department of Materials Science and Metallurgy, University of Cambridge, Cambridge CB3 0FS, UK}

\author{Aron Walsh}
\affiliation{Department of Materials, Imperial College London, Royal School of Mines, Exhibition Road, London, SW7 2AZ, UK}

\date{\today}

\begin{abstract}
Entropy is a critical, but often overlooked, factor in determining the relative stabilities of crystal phases. The importance of entropy is most pronounced in softer materials, where  small changes in free energy can drive phase transitions, which has recently been demonstrated in the case of organic-inorganic hybrid-formate perovskites. In this study we demonstrate the interplay between composition and crystal-structure that is responsible for the particularly pronounced role of entropy in determining polymorphism in hybrid organic-inorganic materials. Using \textit{ab initio} based lattice dynamics we probe the origins and effects of vibrational entropy of four archetype perovskite (ABX$_3$) structures. 
We consider a fully inorganic material (SrTiO$_3$), an A-site hybrid halide material (CH$_3$NH$_3$PbI$_3$), a C-site  hybrid material (KSr(BH$_4$)$_3$) and
   a mixed A- and X-site hybrid-formate material (N$_2$H$_5$Zn(HCO$_2$)$_3$), comparing the differences in entropy between two common
    polymorphs. The results demonstrate the importance of low-frequency inter-molecular modes in determining phase stability in these materials. The understanding gained allows us to propose a general principle for the relative stability of different polymorphs of hybrid materials as temperature is increased.  
\end{abstract}


\maketitle 
The interplay between structure and function is the cornerstone of modern materials design. The ability to rationally predict and
 control how composition affects structure and how, together, they determine materials properties is at the heart of this pursuit.
  Since the early 20$^{\textrm{th}}$ century scientists have developed a host of models to describe how simple elemental properties determine
   crystal structure. The pioneering work of Goldschmidt -- predicting perovskite structures based on radius ratios \cite{Goldschmidt1937} -- and Pauling --
    elucidating the roles of bond valence and electronegativity \cite{Pauling1929} -- stand as some of the most important works in the field of materials design. Other notable successes include the Goodenough-Kanamori rules,\cite{Goodenough1958,Kanamori1959} the hard-soft
  acid/base classification\cite{Pearson1963} and intermetallic alloy formation maps.\cite{Pettifor2003} Recently density functional theory (DFT) has become a powerful tool for assisting materials design.\cite{Garrity2014,Benedek2011,Rondinelli2008}
The rise of quantitatively predictive computational techniques has lead to a proliferation of data --
 the Materials Project \cite{Jain2013a} now contains accurate calculated
  data on over 60,000 known materials. With this wealth of data, the role of simple rules and descriptors to relate
   composition to function are more important than ever. \cite{Ghiringhelli2015} 
  
  \begin{figure}[ht]
\begin{center}
\resizebox{\columnwidth}{!}{\includegraphics*{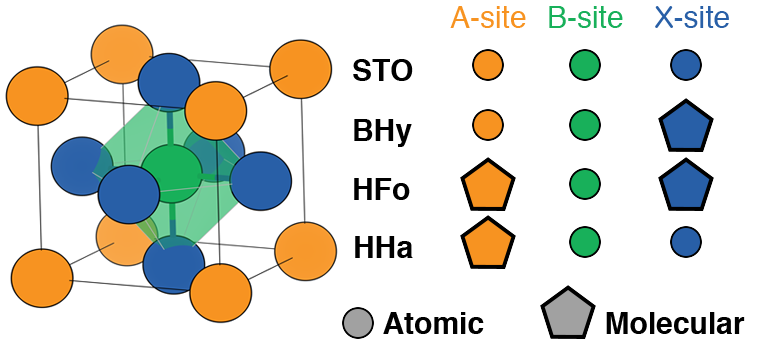}}
\caption{\label{f0} The crystal composition of the perovskite materials with various molecular components studied, demonstrating different types of hybrids. STO (SrTiO$_3$), BHy (KSr(BH$_4$)$_3$), HFo (N$_2$H$_5$Zn(HCO$_2$)$_3$) and HHa (CH$_3$NH$_3$PbI$_3$).} 
\end{center}
\end{figure}
  
The success stories outlined above have all been achieved in the field of purely inorganic materials but today the spectre of hybrid
 organic-inorganic systems looms large. These are materials with a well-defined crystal structure, where at least one Wyckoff site
  is occupied by a molecular species. These materials rose to prominence driven largely by the success of metal organic frameworks (MOFs).\cite{Li1999,Yildirim2005}
    In recent years they have become increasingly of interest in fields such as multi-ferroics\cite{Thomson2012,Stroppa2013} and electronics;\cite{Butler2014}
    notably, the hybrid-perovskite materials have revolutionised the field of photovoltaics.\cite{Frost2014,Zhou542,Li2016}
    
  Despite numerous studies of hybrid materials' properties,\cite{Horcajada2009,Redel2013,Li2013,Svane2015} design principles, 
  such as those reported for inorganic materials, are few and far between.\cite{Kieslich2014,Travis2016} 
   Generally in inorganic solids, the favourable polymorph is that which is enthalpically most stable; statistical and entropic effects are recognised as a major factor other phase transitions, such as those in SrTiO$_3$,\cite{Mller1991} and BaTiO$_3$.\cite{Rabe1997,Stern2004,Qi2016,Senn2016} The role of rotational entropy has been recognised in hybrid perovskite materials;\cite{Duncan2016,Onoda1992} based on accessible orientational microstates  rotational entropy in the region of $R\ln(3)$ - $R\ln(8)$, corresponding to 5.8 - 17.3 $JK^{-1}mol^{-1}$, is predicted.\cite{Duncan2016,Onoda1992} Lattice dynamics have been applied to demonstrate how vibrational entropy can have a critical role to 
   play in determining the stable polymorph of hybrid materials\cite{Kieslich2015,walker2010} and the role of 
   molecule-cavity interactions, determining the mechanical properties of the formate perovskites, has also been recognised.\cite{Kieslich2016}
   Recent computational studies have highlighted the role of entropy in determining crystal polymorph preference in organic crystals\cite{Reilly2014,Vela2014,Nyman2015}. Nonetheless, the role of vibrational entropy in hybrid organic-inorganic materials has generally been underestimated or neglected when predicting the relative stability of polymorphs.\cite{Butler2016cs} 

In this study we probe the basis of the vibrational entropy, which can drive polymorphism in hybrid perovskite materials. We study the materials pictured schematically in Figure ~\ref{f0}, comprised of frameworks of corner-sharing BX$_3$ octahedral units, with A-sites occupying the cavity. We consider SrTiO$_3$ (STO), (CH$_3$NH$_3$)PbI$_3$ (Hyb-Hal), KSr(BH$_4$)$_3$; (Mol-BH$_4$) and
 (N$_2$H$_5$)Zn(HCO$_2$)$_3$ (Hyb-For), comparing the differences in entropy between two common
    polymorphs. STO is studied in its room temperature (cubic) and low temperature (tetragonal) structures. Hyb-Hal is considered in its high temperature (pseudo-cubic) and room temperature (tetragonal) phases\cite{Weller2015} - these vibrational spectra were calculated as part of a previous study.\cite{Brivio2015}.
    Hyb-For is studied in its experimentally reported polymorphs, perovskite (Pna2$_1$) and the channel structure (P2$_1$2$_1$2$_1$).\cite{Kieslich2015} 
     Mol-BH$_4$ is studied in the experimentally determined orthorhombic structure,\cite{Møller2016} which we have also found to be the most stable phase, and a pseudo-cubic structure.
      

\begin{figure}[ht!]
\begin{center}
\resizebox{\columnwidth}{!}{\includegraphics*{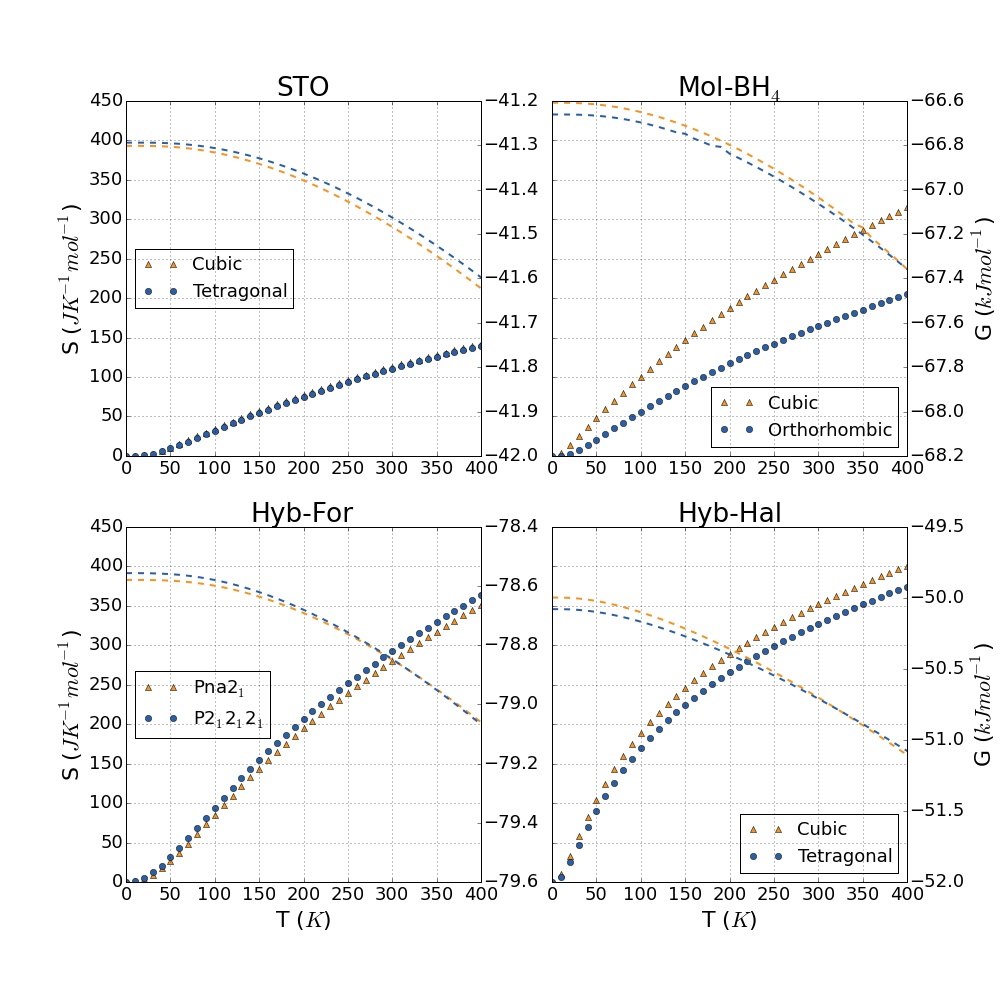}}
\caption{\label{f1} The vibrational entropy (calculated from Equation 1; markers) and Gibbs free energy (dashed lines) for two phases of each of the materials studied across a temperature range. Calculations were performed within the quasi-harmonic approximation.} 
\end{center}
\end{figure}

The free energy of the system is calculated within the quasi-harmonic approximation from 

\begin{equation}
G(T,p) = min_V[U(V) + F_{ph}(T;V) + pV]
\end{equation}
where the the internal energy of the system as a function of volume $U(V)$, is obtained  from DFT calculations. The phonon free energy ($F_{ph}$) arising from harmonic vibrations of the lattice is calculated within the frozen phonon approximation. The effect of thermal expansion of the lattice is approximated by considering the energy-volume relationship. The underlying vibrational entropy is calculated from the phonon density of states of all positive phonon modes according to :

\begin{equation}
\begin{split}
& S_{vib}(T) =  3k_B
\\
& \int_0^{\infty} { g(\varepsilon)[(n(\varepsilon)  + 1) \ln ( n(\varepsilon) + 1) -
  n(\varepsilon) \ln (n(\varepsilon))] d\varepsilon}
\end{split}
\end{equation}

where $k_B$ is the Boltzmann constant, $g(\varepsilon)$ is the normalised phonon density of states with energy $\varepsilon$, $n(\varepsilon)$ is the Bose-Einstein population of a state of energy $\varepsilon$ at temperature $T$ and $\varepsilon = \hbar\omega$, $\omega$ is the mode frequency. The energies and forces of the systems are calculated from DFT, using the PBESol functional\cite{Perdew2008} in the VASP package\cite{Kresse1993} within the projector augmented wave formalism.\cite{Blochl1994} We use a cutoff energy of 500 eV and a $k$-point mesh sampling density with a
target length cut-off of 25 \AA, as prescribed by Moreno and Soler\cite{Moreno1992}. The dynamical matrix and phonon frequencies are obtained using the PHONOPY package.\cite{Togo2015} 

The difference in vibrational entropy between the polymorphs of STO is significantly lower than in the hybrid systems, Figure ~\ref{f1}. This implies that vibrational entropy plays little role in determining the relative stabilities of different phases of this purely inorganic perovskite in the temperature range considered here; configurational entropy is generally more relevant in these materials.


\begin{figure*}[ht]
\begin{center}
{\includegraphics*[width=16.5cm]{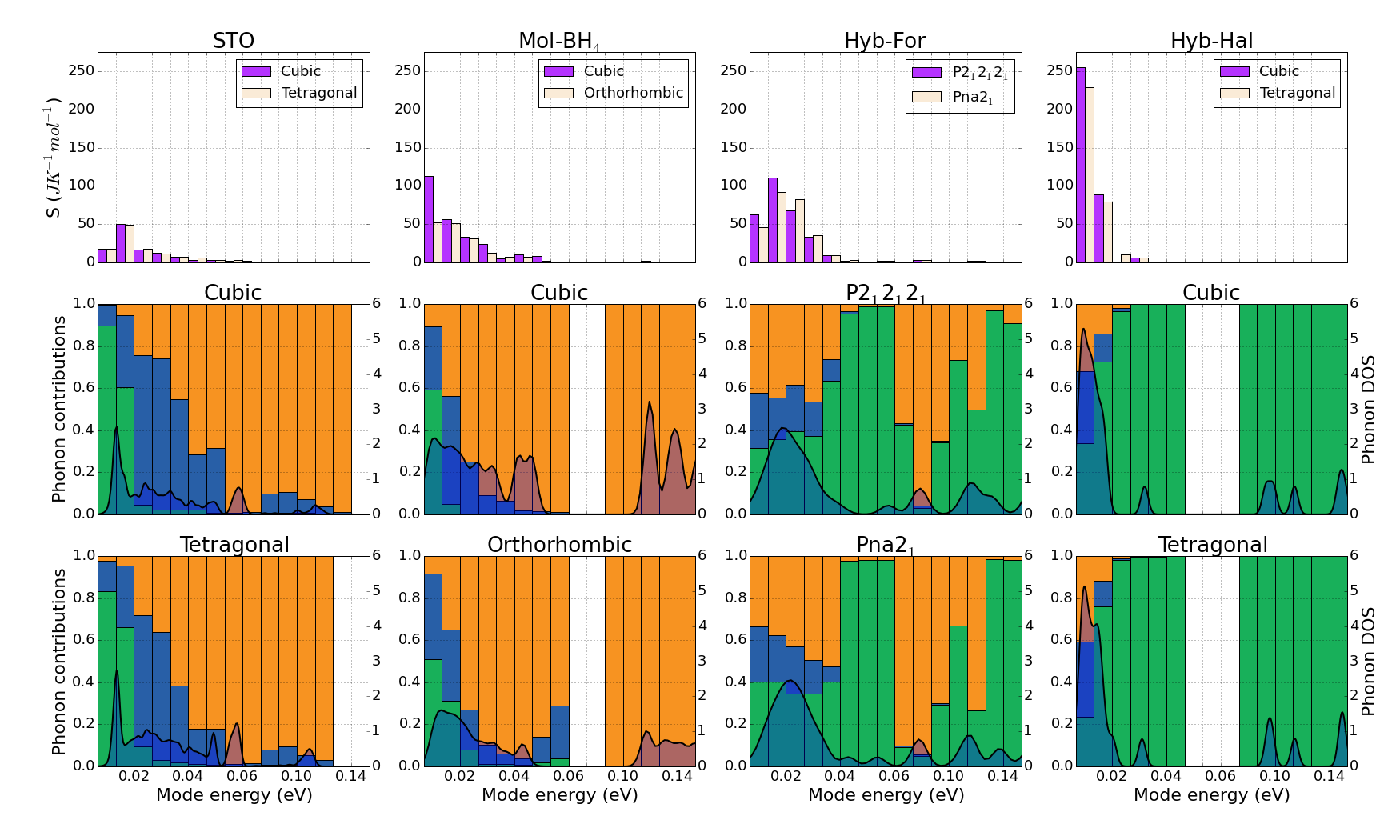}}
\caption{\label{f3} Decomposition of contributions to vibrational entropy for the four compositions studied. Each column is plotted on a common abscissa, where the phonon modes are separated into bins of 0.1 eV. Upper row: The vibrational entropy of both phases decomposed by energy; each 0.1 interval features the entropy of both of the phases studies. Middle and lower rows: The fractional contribution of each crystal site to the total vibrational entropy within the corresponding energy range; A-site (green), B-site(blue), X-site(yellow). The phonon density of states for each polymorph is superimposed on the contribution plot.} 
\end{center}
\end{figure*}

The vibrational entropies of the two polymorphs of all three hybrid materials evolve differently with temperature, resulting in significant differences in $S_{vib}$ at 300 K - $\Delta S$ of the order of 10 - 100 $JK^{-1}mol^{-1}$; at least as large as the rotational entropy and sufficient to shift equilibrium in synthesis of formate perovskites.\cite{Kieslich2015} In all three hybrid materials studied here we find that the order of thermodynamic stability of the polymorphs is reversed as the temperature is increased, Figure~\ref{f1}.
The presence of a discrete molecular unit at the A- or X-site is an important factor in determining differences in vibrational entropies between phases of hybrid-perovskite materials. In order to further analyse this, we deconvolute the phonon contributions to vibrational entropy.


The contributions of the low frequency phonon modes to vibrational entropy of the different phases are compared in the top row of Figure ~\ref{f3} (in the electronic supplementary information we demonstrate that only modes up to 0.04 eV contribute significantly to vibrational entropy at the temperatures considered).
The STO vibrational entropy contribution is significantly lower than found in the other materials. In all three hybrid materials there are differences between the vibrational entropies in the
  various regions of the phonon density of states, but most notably at the low frequency end of the spectrum. This explains why the hybrid materials have vibrational entropy differences between polymorphs large enough to determine relative phase stability close to  room temperature, whilst STO does not.  

 To probe the atomistic origins of the above observation we have decomposed the contributions to the phonon 
 spectrum by crystal site, Figure ~\ref{f3}. We calculate $\langle g_i (\varepsilon)\rangle$, the weighted average of the phonon density of states
  of a given site, $i$, at energy $\varepsilon$:
\begin{equation}
\langle g_i (\varepsilon)\rangle = \frac{\sum_j w_{ij}}{\sum_k w_k}
\end{equation}  
where $w_{ij}$ are the weights of the phonon modes ($w_j$) involving the site $i$ at energy $\varepsilon$ and $w_k$ are all phonon
 mode weights at energy $\varepsilon$ -- we note that the phonon mode weights account for the mass of the ions involved in the modes.

In STO the decomposed DOS are almost identical for each phase; consistent with the finding (Figure~\ref{f1}) that 
 the vibrational entropies are indistinguishable in this material.
In all three hybrid materials the lower end of the phonon spectrum is evenly distributed between the A-site and the framework; indicative of collective
 the phonon modes, which arise from interactions between the framework and the A-site. In the hybrid materials the differences in vibrational entropy between polymorphs, which can determine the balance of thermodynamic stability, are determined by changes in the vibrations in this lower part of the spectrum. 

The smaller number of states at low frequency in STO corresponds to the absence of soft inter-molecular bonding in the inorganic material, where the majority of the 
 bonding is of the strong ionic or covalent type. These bonds are typically much stiffer than inter-molecular bonds and have higher
  vibrational frequencies. 
 In the hybrid systems, each material has a range of soft inter-molecular forces, notably hydrogen bonds, which result in a high DOS
  at the low end of the phonon energy spectrum. 
  
In order to provide a more quantitative measure of the effects of chemical bonding, the degree of vibrational entropy associated with an atomic site can conveniently be expressed by the thermal ellipsoid or ADP (atomic displacement parameter). The ADP provides a Boltzmann averaged measure of the vibrational disorder. The volume of the ellipsoid is determined by the probability of finding the electronic density associated with a given site within the boundaries; it is common to apply a probability of 50 \% . In order to compare the atomic site contributions to the vibrational entropy we can therefore compare the ADPs (here the B value), which are obtained from the phonon spectrum according to :

\begin{equation}
B(j) = \frac{\hbar}{2Nm_j}\sum_{q,v}\omega_v(q)^{-1}(1+2n_v(q))e_v(j,q)\times e_v^{*}(j,q)
\end{equation}
where $N$ is the number of unit cells, $m_j$ is the atomic mass, $\omega_v(q)$ is the frequency, $e_v$ is the polarisation vector and $n_v$ is the thermal population of the band.

 \begin{figure}[ht!]
\begin{center}
\resizebox{\columnwidth}{!}{\includegraphics*{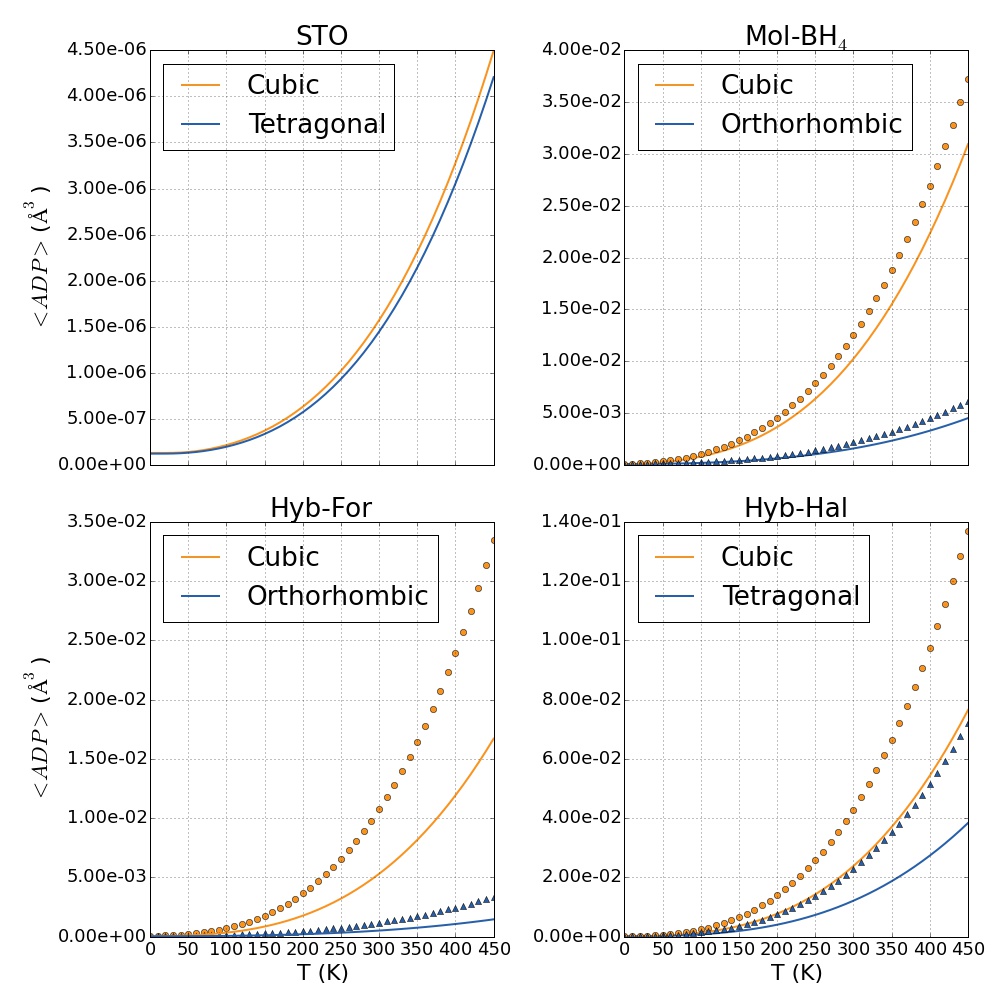}}
\caption{\label{f4} The evolution of atomic displacement volumes with temperature. The anisotropic factor $B$, calculated as in Equation 4, is used to calculate the volume, which is plotted for each polymorph across the temperature range 0 - 450 K. Total mean ADP volumes for all atoms are represented by a line, hydrogen only mean ADP volumes are represented by markers.} 
\end{center}
\end{figure}

The evolution of the mean ADPs with temperature is plotted in Figure~\ref{f4} together with the evolution of the mean volume of the hydrogen ADPs (where they are present). The plot shows that in cases where there is a significant difference between the mean ADP of different polymorphs, the difference arises from the hydrogen based ADPs. In STO, where there are no hydrogens, the ADPs of both polymorphs evolve identically. In all of the hybrid materials the mean ADPs evolve in the same way as the vibrational entropy and the difference between ADPs of the polymorphs is due, almost entirely, to differences between the hydrogen ADPs.

 In all three hybrid materials the higher temperature polymorph is the one with the larger cavity size, either cubic or hexagonal channels (in the Hyb-For). The high-temperature polymorph can thereby maximise the number of weak inter-molecular forces (predominantly H-bonds) and hence increase
 the low frequency DOS. 
 Hyb-For demonstrates this subtle interplay of enthalpy and entropy exquisitely. In the Pna$2_1$  phase the phonon
 DOS has a peak at higher frequency than the P$2_12_12_1$ phase. The higher frequency peak reflects slightly
  stiffer interatomic bonding, related to salt-bridge like interactions between the molecular A-site cation NH$_3$NH$_2^+$ and the pseudocubic cavity. The total free energy also contains a contribution from vibrational entropy. As the temperature rises the latter contribution begins to outweigh the 
former and the phase with softer more flexible bonding becomes preferred. We note, the importance of weak hydrogen-bonding interactions for determining the stability of competing polymorphs has previously been observed experimentally in organic crystal systems and in lithium tartrate crystals. \cite{Yeung2014}

In conclusion, we have studied the origins of vibrational entropy differences in hybrid-perovskite materials, by applying lattice dynamics calculations based on quantum mechanical forces.   
 Our analysis allows us to quantify the contribution of vibrational entropy to total energies and demonstrate its crucial role in determining the stable polymorph; providing a theoretical basis for the general principle that in hybrid organic-inorganic materials the materials favour structures that maximise the number of soft inter-molecular interactions as the temperature rises. The findings suggest that, in the growing class of hybrid-perovskite materials, vibrational contributions from amine cage interactions will be crucial for determining structure and properties. The contribution from vibrational entropy is an important consideration for the development of composition-structure-property relationships in the emerging field of hybrid materials.

We acknowledge membership of the UK's HPC Materials Chemistry Consortium (EPSRC EP/L000202) and access to computational resources through PRACE.
A.W. acknowledges support from the Royal Society for a University Research Fellowship and K.T.B. is funded by EPSRC (EP/M009580/1 and EP/J017361/1).
GK is the holder of a
DFG fellowship (KI1870). GK and AKC gratefully thank the
Ras Al Khaimah Center for Advanced
Materials for financial support.

Data analysis scripts used to generate Figures 2-4, the optimised structures, and
data from the phonon calculations are available online, free
of charge, from https://github.com/WMD-group/Phonons

\bibliography{Lib1.bib}
\end{document}